\documentclass[oneside,english]{amsart}

\usepackage{cmap}

\usepackage[a4paper,includeheadfoot,margin=3.64cm]{geometry}

\usepackage{beton}
\usepackage{euler}
\usepackage[T1]{fontenc}
\usepackage[latin9]{inputenc}
\usepackage{varioref}
\usepackage{amsthm}
\usepackage{amssymb}

\usepackage[usenames,dvipsnames]{xcolor}

\usepackage{hyperref}
\hypersetup{colorlinks=true,linkcolor=MidnightBlue,backref=page,citecolor=Magenta}

\usepackage[backend=bibtex,style=alphabetic,firstinits=true,doi=false]{biblatex}
\usepackage{mathtools}

\usepackage[shortlabels]{enumitem}

\makeatletter
\numberwithin{equation}{section} %
\numberwithin{figure}{section} %
\theoremstyle{plain}
\theoremstyle{plain}
\newtheorem{theorem}{Theorem}
  \theoremstyle{plain}
  \newtheorem{lemma}{Lemma}
  \theoremstyle{plain}
  \newtheorem{proposition}{Proposition}
  \theoremstyle{remark}
  \newtheorem*{note*}{Note}
  \theoremstyle{remark}
  \newtheorem*{conclusion*}{Conclusion}
  \theoremstyle{remark}
  \newtheorem{remark}{Remark}
 \theoremstyle{definition}
  
  \newtheorem{definition}{Definition}
  \theoremstyle{plain}
  \newtheorem{corollary}{Corollary}

\usepackage{amsthm}
\usepackage{amsfonts}
\usepackage{amscd}
\usepackage[all]{xy}
\usepackage{pb-diagram,pb-xy}

\usepackage{tikz}
\usetikzlibrary{matrix,arrows}
\usetikzlibrary{shapes.geometric,fit}

\tikzstyle{legend_general}=[rectangle, rounded corners, thin,
                          top color= white,bottom color=lavander!25,
                          minimum width=2.5cm, minimum height=0.8cm,
                          violet]

\tikzset{
   dashellipse/.style={rectangle,draw,dashed,inner sep=8pt,fit={#1}}
}

\usepackage{tikz-cd}
\usetikzlibrary{decorations.pathreplacing, positioning}

\usepackage{soul}

\newcommand{\g}{\mathfrak{g}}

\newcommand{\kf}{\mathfrak{k}}
\newcommand{\hf}{\mathfrak{h}}
\newcommand{\pf}{\mathfrak{p}}

\newcommand{\cF}{{\mathcal F}}
\newcommand{\cV}{{\mathcal V}}

\newcommand{\cI}{{\mathcal I}}

\newcommand{\cL}{{\mathcal L}}

\newcommand{\mR}{\mathbb{R}}
\newcommand{\mC}{\mathbb{C}}

\usepackage[e]{esvect}

\usepackage{charter}

\makeatother

\makeatletter
\newcommand*\bigcdot{\mathpalette\bigcdot@{.5}}
\newcommand*\bigcdot@[2]{\mathbin{\vcenter{\hbox{\scalebox{#2}{$\m@th#1\bullet$}}}}}
\makeatother

\usepackage{babel}

\begin{document}

\title[Geometrical setting for gravity and Chern-Simons]{Geometrical setting for the correspondence between gravity and Chern-Simons field theories}

\author{S. Capriotti}
\address{Departamento de Matem\'atica \\
Universidad Nacional del Sur and Conicet\\
 8000 Bah{\'\i}a Blanca \\
Argentina}
\email{santiago.capriotti@uns.edu.ar}

\maketitle

\begin{abstract}
  In the present article, Chern-Simons gauge theory and its relationship with gravity are revisited from a geometrical viewpoint. In this setting, our goals are twofold: In one hand, to show how to represent the family of variational problems for Chern-Simons theory in the set of $K$-structures on the space-time $M$, by means of a unique variational problem of Griffiths type. On the other hand, to make an interpretation of the correspondence of these theories in terms of the correspondence between the frame bundle and its affine counterpart.
\end{abstract}

\section{Integrable field theories and gravity}
\label{sec:chain-integr-field}

Chern-Simons field theory is a well-known type of gauge field theory \cite{10.2307/1971013,PhysRevLett.48.975,Freed:1992vw} whose quantization yields to topological field theory \cite{Witten1988,Ramadas1989}. In its more general setting \cite{Freed:1992vw}, the fields in Chern-Simons gauge theory are connections on any $K$-principal bundle with fixed base space. In this vein, the Chern-Simons action is considered as a function on the (infinite dimensional) manifold of the $K$-connections on a fixed base space $M$, and it is evident that the construction of this manifold requires a precise knowledge of every $K$-principal bundle on $M$. Although this operation can be performed successfully (getting more complicated as the dimension of $M$ increases), this scheme is out of range of the geometrical formulation for field theory \cite{Gotay:1997eg,Blee,doi:10.1142/9693,deLen2003,PROP:PROP2190440304}, because in this approach the fields should be sections of a definite bundle (preferably, of finite dimension). For example, in \cite{TejeroPrieto2004} is described a scheme for Chern-Simons field theory fitting in these geometrical requirements, based in a variational problem posed by local Lagrangians. Another mathematical formulation in this vein, using Cartan connections as fields, was presented in \cite{wise2009symmetric}.

\par On the other hand, it is an interesting result \cite{ACHUCARRO198689,MR974271} that when gauge group is the Poincar\'e group $\text{ISO}\left(2,1\right)$, Chern-Simon field theory in dimension $3$ can be related with Palatini gravity on a spacetime of the same dimension. This correspondence is achieved by the splitting
\[
  \mathfrak{iso}\left(2,1\right)=\mathfrak{so}\left(2,1\right)\oplus\mR^3,
\]
that decomposes the field $A$ in two parts, one living in $\mR^3$ and another in $\mathfrak{so}\left(2,1\right)$; the idea is to recognize each of these fields as a vierbein and a $\mathfrak{so}\left(2,1\right)$-connection respectively, which can be seen as the basic fields for the Palatini description of general relativity. In \cite{wise2009symmetric}, this scheme is generalized to pair of algebras $\g\subset\hf$ defining a Cartan connection on a $G$-principal bundle $\pi:P\to M$.

\par A remarkable fact is that, in some of these descriptions of Chern-Simons theory and its connection with gravity, is usual a certain lack of definiteness with respect to the nature of the principal bundle to which the connections of the theory belongs. While the corresponding base space and structure group are properly set, it is avoided any precision on the geometrical characteristics of such bundle. Not that it keeps people away from working with Chern-Simons gauge theory: It is in this context where frameworks like the one discussed by Freed are fruitful. Moreover, the informed reader could recall that when dealing with gravity, a metric is available, and it can be used to select suitable subbundles of the frame bundle. Nevertheless, this answer should be considered as partial, and in fact might put us in a paradoxical situation, because the metric is part of the dynamical fields in gravity, and so it should be necessary to solve the equations of motion of gravity before constructing the bundle where the fields should live. Therefore, it seems to appear a tension between the formalism describing field theory from a geometric viewpoint, and the characteristics that a Chern-Simons gauge theory must have in order to represent gravity. At the end, the apparent paradox is solved by invoking the gauge symmetry of the Lagrangian, but it could be interesting to explore how to deal with this paradox from a geometrical point of view.

\par As we mentioned above, when working with Chern-Simons theory from the viewpoint of geometric mechanics, the extremals of the variational problem describing it should live on the space of sections of a unique bundle (preferably finite dimensional). On the other hand, although the structure group for the principal bundle used for the usual Palatini gravity description is the Lorentz group \cite{Hehl19951,0264-9381-22-19-016}, there is in general no canonical principal bundle with this group as structure group. In fact, recall the bundle of frames
\[
  \tau:LM\to M
\]
with structure group $GL\left(m,\mR\right)$ (see Definition \eqref{eq:FrameBundle} below). If the extremals of Chern-Simons gauge theory are determined in some extent from sections of a connection bundle with structure group $K\subset GL\left(m,\mR\right)$, it would be nice to find the associated principal bundle between the $K$-structures on $M$, that is, between the $K$-principal subbundles of the frame bundle. But, when $K=SO\left(p,q\right),p+q=\dim{M}$, it is equivalent to have a section of the quotient bundle
\[
  \tau_\Sigma:\Sigma:=LM/K\to M,
\]
namely, to select a metric on $M$ with the given signature.

\par These considerations set the aims of the article. In first place, we are seeking to find a kind of universal variational problem, whose extremals represent the extremals of any Chern-Simons field theory formulated on the $K$-structures of $M$. It should be noted that it solves the paradoxical situation mentioned above, because the space of extremals of the variational problems on any of these $K$-structures is equivalent to the space of extremals of the universal variational problem. Finally, we want to use this general setting in order to describe the map that relates the sections of these $K$-structures with the sections of a formulation of Palatini gravity. The main result in this direction is the following: we can construct the bundle of affine frames $\beta:AM\to LM$, essentially by enlarging the group structure from $GL\left(m,\mR\right)$ to
\[
  A\left(m,\mR\right)=GL\left(m,\mR\right)\ltimes\mR^m,
\]
the group of affine transformation of $\mR^m$. This construction extends to the bundles of connections, and in fact, there exists a one-to-one correspondence between the connections associated to every bundle, if we fix the $\mR^m$-part by using the canonical solder form $\theta\in\Omega^1\left(LM,\mR^m\right)$. In this setting, the other interesting result achieved in the present work tells us that Palatini gravity, considered as variational problem on the jet space of the frame bundle $J^1\tau$, can be considered as equivalent to the universal Chern-Simons gauge theory on $J^1\left(\tau\circ\beta\right)$, the jet space of the affine frame bundle on $M$. Namely, the universal variational problem for Chern-Simons theory is nothing but a variational problem for Palatini gravity, but instead lifted to the affine frame bundle using a classical construction \cite{KN1}.

\par Let us briefly describe the structure of the article. The geometrical tools used throughout the article are presented in Section \ref{sec:some-geom-tools}. Given a $K$-principal bundle, the collection of local variational problems on the associated bundle of connections described in \cite{TejeroPrieto2004} are related to a global variational problem in Section \ref{sec:local-glob-vari}; in order to deal with this comparison, a special kind of variational problems is introduced in this section. The main result of the section is Theorem \ref{thm:Local-to-global}. The universal variational problem for Chern-Simons gauge is constructed in Section \ref{sec:famil-princ-subb}; Theorem \ref{thm:univ-chern-simons} proves that locally, its extremals are in one-to-one correspondence with the extremals of Chern-Simons theory on any $K$-subbundle, provided that some condition on this family holds. The classical construction of the affine frame bundle is reviewed in Section \ref{sec:lift-conn-affine}. Chern-Simons gauge variational problem for Lie group $A\left(3,\mR\right)$ is set in Section \ref{sec:chern-simons-vari}; it is achieved by specializing the universal variational problem defined in Section \ref{sec:famil-princ-subb} to the particular case where the gauge group is the general affine group $A\left(m,\mR\right)$, and some particularities are discussed regarding the invariant bilinear form when $m=3$. The relation between this universal variational problem for Chern-Simons theory and Palatini gravity is considered in Section \ref{sec:gauge-theory-gravity}; the main result of the section becomes Corollary \ref{thm:chern-simons-palatini}, where it is proved the equivalence between this universal variational problem on the jet bundle of the affine frame bundle and the description of Palatini gravity as a variational problem on the jet bundle of the (usual) bundle of frames.  

\subsubsection*{Notations} We are adopting here the notational conventions from \cite{saunders89:_geomet_jet_bundl} when dealing with bundles and its associated jet spaces. It means that, given a bundle $\pi:P\to M$, there exists a family of bundles and maps fitting in the following diagram
\[
  \begin{tikzcd}[ampersand replacement=\&,row sep=.5cm,column sep=1cm]
    \cdots
    \arrow{r}{}
    \&
    J^{k+1}\pi
    \arrow{r}{\pi_{k+1,k}}
    \arrow[swap]{rrdd}{\pi_{k+1}}
    \&
    J^{k}\pi
    \arrow{r}{\pi_{k,k-1}}
    \arrow{rdd}{\pi_{k}}
    \&
    \cdots
    \arrow{r}{\pi_{21}}
    \&
    J^1\pi
    \arrow[swap]{ddl}{\pi_1}
    \arrow{r}{\pi_{10}}
    \&
    P
    \arrow{lldd}{\pi}
    \\
    \&
    \cdots
    \&
    \&
    \cdots
    \&
    \&
    \\
    \&
    \&
    \&
    M
    \&
    \&
  \end{tikzcd}  
\]
Sections of $\pi:P\to M$ will be indicate by the symbol $\Gamma\pi$. The set of vectors tangent to $P$ in the kernel of $T\pi$ will be represented with the symbol $V\pi\subset TP$. In this regard, the set of vector fields which are vertical for a bundle map $\pi:P\to M$ will be indicated by $\mathfrak{X}^{V\pi}\left(P\right)$. The space of differential $p$-forms, sections of $\Lambda^p (T^*Q)\to Q$, will be denoted by $\Omega^p(Q)$. {We also write $\Lambda^\bullet(Q)=\bigoplus_{j=1}^{\dim Q}\Lambda^j(T^*Q)$}. If $f\colon P\to Q$ is a smooth map and $\alpha_x$ is a $p$-covector on $Q$, we will sometimes use the notation $\alpha_{f(x)}\circ T_xf$ to denote its pullback $f^*\alpha_x$. If $P_1\to Q$ and $P_2\to Q$ are fiber bundles over the same base $Q$ we will write $P_1\times_Q P_2$ for their fibered product, or simply $P_1\times P_2$ if there is no risk of confusion. Unless explicitly stated, the canonical projections onto its factor will be indicated by
\[
  \text{pr}_i:P_1\times P_2\to P_i,\qquad i=1,2.
\]
Given a manifold $N$ and a Lie group $G$ acting on $N$, the symbol $\left[n\right]_G$ for $n\in N$ will indicate the $G$-orbit in $N$ containing $n$; the canonical projection onto its quotient will be denoted by
\[
  p_G^N:N\to N/G.
\]
Also, if $\mathfrak{g}$ is the Lie algebra for the group $G$, the symbol $\xi_N$ will represent the infinitesimal generator for the $G$-action asssociated to $\xi\in\mathfrak{g}$. Finally, Einstein summation convention will be used everywhere.

\section{Some geometrical tools}
\label{sec:some-geom-tools}

Throughout the article, we will make extensive use of the geometrical tools related to the jet space associated to a principal bundle, as well as its connection bundle, as they are discussed in \cite{MR0315624,springerlink:10.1007/PL00004852}. So, in order to proceed, let $p:Q\to N$ be a principal bundle with structure group $H$; then we can lift the right action of $H$ on $Q$ to a $H$-action on $J^1p$, and so define the bundle
\[
  \overline{p}:C\left(Q\right):=J^1p/H\to M
\]
fitting in the following diagram
\[
  \begin{tikzcd}[ampersand replacement=\&,row sep=.7cm,column sep=.7cm]
    \&
    J^1p
    \arrow[swap]{dl}{p_{10}}
    \arrow{dr}{p_H^{J^1p}}
    \&
    \\
    Q
    \arrow[swap]{dr}{p}
    \&
    \&
    C\left(Q\right)
    \arrow{dl}{\overline{p}}
    \\
    \&
    M
    \&
  \end{tikzcd}  
\]
It can be proved that this diagram defines $J^1p$ as a pullback, namely, that
\[
  J^1p=p^*Q=Q\times_M C\left(Q\right).
\]
We know that $J^1p$ comes equipped with the \emph{contact structure}, that can be described by means of a $Vp$-valued $1$-form
\[
  \left.\theta\right|_{j_x^1s}:=T_{j_x^1s}p_{10}-T_xs\circ T_{j_x^1s}p_1;
\]
moreover, because $p:Q\to N$ is a principal bundle, we have the bundle isomorphism on $Q$
\[
  Vp\simeq Q\times\hf.
\]
It means that we can consider $\theta$ as a $\hf$-valued $1$-form; in fact, with respect to the $K$-principal bundle structure
\[
  p_H^{J^1p}:J^1p\to C\left(Q\right),
\]
the $1$-form $\theta$ becomes a connection form, dubbed \emph{canonical connection form}. It has the following property.

\begin{proposition}
  Let $\Gamma_Q:Q\to J^1p$ be a connection on $Q$. Then its connection form $\omega_Q\in\Omega^1\left(Q,\hf\right)$ can be constructed from the canonical connection form through pullback along $\Gamma_Q$,
  \[
    \omega_Q=\Gamma_Q^*\theta.
  \]
\end{proposition}

For any manifold $M$ of dimension $m$, the bundle $\tau:LM\to M$ is the \emph{frame bundle of $M$}, is defined through
\begin{equation}\label{eq:FrameBundle}
  LM:=\bigcup_{x\in M}\left\{u:\mR^m\to T_xM\text{ linear and with inverse}\right\}.
\end{equation}
It has a canonical free $GL\left(m,\mR\right)$-action, given by the formula 
\[
  u\cdot A:=u\circ A,\qquad u\in LM,A\in GL\left(m,\mR\right).
\]

Let $\eta\in M_m\left(\mR\right)$ be a non degenerate symmetric $m\times m$-matrix with real entries; for definiteness, we will fix
\[
  \eta:=
  \begin{bmatrix}
    -1&0&\cdots&0\\
    0&1&&0\\
    \vdots&&\ddots&\vdots\\
    0&\cdots&0&1
  \end{bmatrix},
\]
although the constructions we will consider in the present article should work with any signature. Then we have a Lie group $K\subset GL\left(m,\mR\right)$ defined by
\begin{equation}\label{eq:LorentzGroup}
  K:=\left\{A\in M_m\left(\mR\right):A\eta A^T=\eta\right\}.
\end{equation}
Then we have an action of $K$ on $LM$; it yields to a bundle
\[
  \tau_\Sigma:\Sigma:=LM/K\to M.
\]
\begin{lemma}
  The bundle $\Sigma$ is the bundle of metrics of $\eta$-signature on $M$.
\end{lemma}
Let us indicate with $\kf$ the Lie algebra of $K$; then, we have that
\[
  \kf:=\left\{a\in\mathfrak{gl}\left(m,\mR\right):a\eta+\eta a^T=0\right\}.
\]
Accordingly, we can define the $K$-invariant subspace
\begin{equation}\label{eq:Transvections}
  \pf:=\left\{a\in\mathfrak{gl}\left(m,\mR\right):a\eta-\eta a^T=0\right\},
\end{equation}
usually called \emph{transvections} (see \cite{wise2009symmetric}); it follows that
\[
  \mathfrak{gl}\left(m,\mR\right)=\kf\oplus\pf.
\]

We will need this result concerning some natural properties of the canonical connections.

\begin{lemma}
  Let $p:Q\to N$ be a $H$-principal bundle and $i_\zeta:R_\zeta\hookrightarrow Q$ a $L$-principal subbundle, with projection
  \[
    p_\zeta:R_\zeta\to N.
  \]
  We have the following relation between the canonical connections on $J^1p$ and $J^1p_\zeta$, namely
  \[
    \left(j^1i_\zeta\right)^*\omega=A,
  \]
  where $\omega\in\Omega^1\left(J^1p,\mathfrak{h}\right)$ and $A\in\Omega^1\left(J^1p_\zeta,\mathfrak{l}\right)$ are the corresponding canonical connection forms.
\end{lemma}
\begin{proof}
  The lemma follows from the formula
  \[
    \left.\omega\right|_{j_x^1s}\circ T_{j_x^1\sigma}\left(j^1i_\zeta\right)=T_{\sigma\left(x\right)}i_\zeta\circ\left.A\right|_{j_x^1\sigma},
  \]
  valid for any $j_x^1s=j^1i_\zeta\left(j_x^1\sigma\right)$ and $j_x^1\sigma\in J^1p_\zeta$.
\end{proof}

Assuming that some topological conditions on the manifold $M$ hold\footnote{The existence of a metric with $\left(p,q\right)$-signature is equivalent to the splitting of the tangent bundle $TM$ in a direct sum of vector subbundles of rank $p$ and $q$ respectively.}, a family
\[
  \left\{O_\zeta:\zeta:M\to\Sigma\right\}
\]
of subbundles of $LM$ can be constructed; namely, let us define
\[
  O_\zeta:=\left\{u\in LM:\zeta\left(u\left(w_1\right),u\left(w_2\right)\right)=\eta\left(w_1,w_2\right)\text{ for all }w_1,w_2\in\mR^m\right\}.
\]
Here $\eta:\mR^m\times\mR^m\to\mR$ indicates the bilinear form associated to the matrix $\eta$. Because every member of this family is a $K$-reduction for $LM$, it has the property that for every (local) section $\sigma:U\to LM$ there exists a metric $\zeta_0:M\to\Sigma$ such that
\[
  \sigma\left(x\right)\in O_{\zeta_0}
\]
for any $x\in U$. In order to generalize this property, we will introduce the following definition.
\begin{definition}[Complete family of subbundles for a principal bundle]\label{def:complete-family}
Let $p:Q\to N$ be a principal bundle and $\cF:=\left\{p_\zeta:R_\zeta\to N\right\}$ a family of subbundles of $Q$ with the following property: For every (local) section $s\in\Gamma p$, there exists $R_\zeta\in\cF$ such that $s\in\Gamma p_\zeta$. We will call to such family a \emph{complete family of subbundles}.
\end{definition}

\section{Local and global variational problems for Chern-Simons field theory}
\label{sec:local-glob-vari}

Our first task is to establish a global variational problem for Chern-Simons gauge theory, and to relate it with the formulation of Chern-Simons variational problem as a collection of local variational problems, as it is done in \cite{TejeroPrieto2004}. 

\subsection{Definitions}
\label{sec:definitions}

In order to proceed, it will be necessary to work with a kind of variational problem slightly more general than those usually considered in field theory \cite{garcia1974poincare,KrupkaVariational}.

\begin{definition}[Variational problem on a bundle]\label{def:griffiths-vars-prob}
  A \emph{variational problem on $E$} is specified by a triple
  \[
    \left(p:E\to N,\lambda,\mathcal{I}\right)
  \]
  where $p:E\to N$ is a bundle over a $n$-dimensional manifold $N$, $\lambda\in\Omega^n\left(E\right)$ is a $n$-form (called \emph{Lagrangian form}) and $\cI\subset\Omega^\bullet\left(E\right)$ is an ideal in the exterior algebra of $E$, closed by exterior differentiation. The \emph{underlying variational problem} consists into finding the extremals of the action
  \[
    \sigma\mapsto\int_N\sigma^*\lambda,
  \]
  where $\sigma\in\Gamma p$ belongs to the set of sections of $p:E\to N$ that are integral for the ideal $\cI$, namely, such that $\sigma^*\alpha=0$
  for every $\alpha\in\cI$.
\end{definition}

\begin{remark}
  Sections $s:N\to A$ that verify the constraints imposed by $\cI$, namely, such that
  \[
    s^*\alpha=0
  \]
  for all $\alpha\in\cI$, are referred to as \emph{admissible sections}.
\end{remark}
\begin{remark}
  A variational problem is \emph{classical} (also called \emph{usual}) when the bundle is the jet space $p_1:J^1p\to N$ of a bundle $p:E\to N$, the Lagrangian form is $p_1$-horizontal, and ideal $\cI$ is the \emph{contact structure $\cI_{\text{con}}$ of $J^1p$}. If $\left(x^i,u^A\right)$ are a set of adapted coordinates on $E$, this ideal is locally generated by the set of forms
  \[
    \cI_{\text{con}}=\left\{du^A-u_i^Adx^i,du^B_k\wedge dx^k\right\}.
  \]
  The equations describing the extremals of a classical variational problem are the so called \emph{Euler-Lagrange equations}.
  Variational problems described in Definition \ref{def:griffiths-vars-prob} are also known in the literature as \emph{Griffiths variational problems} \cite{book:852048,hsu92:_calcul_variat_griff,GotayCartan}.
\end{remark}

There is another generalization when dealing with classical variational problems \cite{TejeroPrieto2004}.

\begin{definition}[Classical variational problems defined by local data]
  Let $\left\{U_\alpha\right\}$ be a open covering of $N$, and define the open sets
  \[
    V_\alpha:=p_1^{-1}\left(U_\alpha\right);
  \]
  let $\cI_{\text{con}}^\alpha$ be the pullback of the contact structure of $J^1p$ to $V_\alpha$ for all $\alpha$. A collection of classical variational problems
  \[
    \mC:=\left\{\left(\left.p_1\right|_{V_\alpha}:V_\alpha\to U_\alpha,\cL_\alpha,\cI_{\text{con}}^\alpha\right)\right\}
  \]
  is a \emph{classical variational problem defined by the local data $\left\{U_\alpha,\cL_\alpha\right\}$} if the Euler-Lagrange equations for its extremals coincide on every intersection $U_\alpha\cap U_\beta$.

  A pair of classical variational problems defined by local data are equivalent if and only if its Euler-Lagrange equations coincide whenever its domains intersect.
\end{definition}

\begin{remark}
  Given a open covering $\left\{U_\alpha\right\}$ for $N$ and a classical variational problem
  \[
    \left(p:J^1p\to N,\cL,\cI_{\text{con}}\right),
  \]
  there exists an equivalent classical variational problem defined by local data, where $\cL_\alpha$ is the pullback to $V_\alpha$ of the global form $\cL$.
\end{remark}

\subsection{Liftings of variational problems}
\label{sec:lift-vari-probl}

To compare variational problems of different nature, it could be interesting to have a way to lift a variational problem from a bundle to its jet bundle.

\begin{definition}[Horizontalization operator]
  Let $k,q$ be positive integers and $p:Q\to N$ a bundle. The \emph{horizontalization operator} is the map
  \[
    h:\Omega^{q}\left(J^kp\right)\to\Omega^{q}\left(J^{k+1}p\right)
  \]
  defined by the formula
  \[
    \left.h\left(\alpha\right)\right|_{j_x^{k+1}s}:=\left.\alpha\right|_{j_x^ks}\circ T_xj^ks\circ T_{j_x^{k+1}s}p_{k+1}
  \]
  for all $j_x^{k+1}s\in J^{k+1}p$.
\end{definition}

For $\left(x^i,u^A\right)$ a set of adapted coordinates on $p:Q\to N$, such that $\left(x^i,u^A,u^A_J\right)$ are the induced coordinates on $J^kp$ ($J$ indicates multiindices), then we have that
\[
  h\left(dx^i\right)=dx^i,\qquad h\left(du^A\right)=u^A_idx^i,\qquad h\left(du^A_J\right)=u^A_{J+i}dx^i.
\]

\begin{definition}[Lifting of a variational problem]
  Let $p:E\to N^n$ be a bundle, $\lambda\in\Omega^n\left(E\right)$ a Lagrangian $n$-form and $\alpha_1,\cdots,\alpha_q\in\Omega^1\left(E\right)$ a set of forms. Let $\mathcal{V}$ be the variational problem with action 
  \[
    s\mapsto\int_Us^*\left(\lambda\right)
  \]
  and constraints $s^*\alpha_i=0$ for all $1\leq n\leq q$. The \emph{lifting of $\mathcal{V}$} is the variational problem $\mathcal{V}^{\left(1\right)}$ on $J^1p$, defined by the Lagrangian $n$-form
  \[
    \lambda^{\left(1\right)}:=h\left(\lambda\right)\in\Omega^n\left(J^1p\right)
  \]
  and with constraints defined by the set of forms $\beta_i:=p_{10}^*\alpha_i$, for $1\leq i\leq q$, and the contact structure.
\end{definition}

The relevance of the lifting of a variational problem is shown by the following result.

\begin{proposition}
  There exists a one-to-one correspondence between the extremals of a variational problem $\mathcal{V}$ and its lifting $\mathcal{V}^{\left(1\right)}$. This correspondence is given by
  \[
    s\mapsto j^1s\qquad\text{and}\qquad\sigma\mapsto p_{10}\circ\sigma.
  \]
\end{proposition}
\begin{proof}
  Let $s:V\to E$ be an extremal for $\cV$; then
  \[
    j^1s:V\to J^1p
  \]
  is an admissible section for the set of constraints $\left\{\beta_i,\theta\text{ contact}\right\}$. Also
  \[
    \int_V\left(j^1s\right)^*h\lambda=\int_Vs^*\lambda;
  \]
  therefore, if $\left\{s_t:V\to E\right\}$ is a curve of admissible sections such that $s_0=s$, then the function
  \[
    f\left(t\right):=\int_V\left(j^1s_t\right)^*h\lambda
  \]
  will have a critical point in $t=0$.

  Conversely, let $\sigma:V\to J^1p$ be an extremal for the variational problem $\cV^{\left(1\right)}$. Then we have that
  \[
    \sigma=j^1\left(p_{10}\circ\sigma\right),\qquad\left(p_{10}\circ\sigma\right)^*\alpha_i=0,
  \]
  and the result follows.
\end{proof}

\subsection{Local and global Chern-Simons Lagrangians}
\label{sec:local-global-chern}

Let $K$ be a Lie group and $\pi_\zeta:R_\zeta\to M$ a $K$-principal bundle (the notation will be explained later); on its first order jet space
\[
  \left(\pi_\zeta\right)_1:J^1\pi_\zeta\to M
\]
we will define a (global) variational problem, which we will prove  to represent Chern-Simons gauge theory. In order to accomplish this task, it will be necessary to lift it to $J^1\left(\pi_{\zeta}\right)_1$, and compare it with the variational problem defined by local data, which lives on $J^1\overline{\pi}_\zeta$.

Let us now suppose that we have an invariant polynomial $q:\kf\to\mR$ of degree $n$. According to the Chern-Simons theory \cite{10.2307/1971013,Morita1}, the $2n$-form
\[
  \alpha:=q\left(F\right)\in\Omega^{2n}\left(J^1\pi_\zeta\right)
\]
is closed.
\begin{remark}
  When the Lie algebra comes with an invariant bilinear form, we can consider the quadratic polynomial
  \[
    q\left(F\right):=\left<F\stackrel{\wedge}{,}F\right>;
  \]
  in this case, the transgression is given by
  \[
    Tq\left(A,F\right)=\left<A\stackrel{\wedge}{,}F\right>-\frac{1}{6}\left<A\stackrel{\wedge}{,}\left[A\stackrel{\wedge}{,}A\right]\right>.
  \]
\end{remark}
In any case, the transgression formula can be applied to the bundle
\[
  \text{pr}_1:J^1\pi_\zeta\times_{C\left( {R}_\zeta\right)} J^1\pi_\zeta\to J^1\pi_\zeta,
\]
which is a trivial $K$-principal bundle; it means that $\alpha$ is not only closed, but also exact. Therefore, there exists a $\left(2n-1\right)$-form
\[
  \beta:=Tq\left(A,F\right)\in\Omega^{2n-1}\left(J^1\pi_\zeta\right),
\]
such that
\[
  d\beta=\text{pr}_1^*\alpha.
\]
On the other hand, we have another $K$-principal bundle structure, namely the quotient map
\[
  p_{K}^{J^1\pi_\zeta}:J^1\pi_\zeta\to C\left( {R}_\zeta\right),
\]
and $q\left(F\right)$ defines a Chern class for it. Accordingly, there exists a $2n$-form $\gamma$ on $C\left( {R}_\zeta\right)$ such that
\[
\left(p_{K}^{J^1\pi_\zeta}\right)^*\gamma=q\left(F\right).
\]
Moreover, using the canonical $2$-form $F_2$ on the bundle of connections $C\left( {R}_\zeta\right)$, we can prove that
\[
  \gamma=q\left(F_2\right).
\]
But now, this bundle is not trivial in general; in short, from decomposition
\[
  J^1\pi_\zeta= {R}_\zeta\times_M C\left( {R}_\zeta\right),
\]
we obtain that it is trivial if and only if the bundle
\[
  \pi_\zeta: {R}_\zeta\to M
\]
is. In consequence, it is not expected that the $2n$-form $\gamma$, although closed, should also be exact; it is the reason why, although we have a global variational problem on $J^1\pi_\zeta$, it cannot be reproduced on $C\left( {R}_\zeta\right)$, even having in mind that the transformations properties of $\cL_{CS}$ are telling us that the degrees of freedom associated to the $ {R}_\zeta$-factor can be ignored.

Nevertheless, because $ {R_\zeta}$ admits trivializing open sets, the previous considerations can be used to associate to every such set $U\subset M$ a Lagrangian $\cL_U:=Tq\left(A_2^U,F_2\right)$, where
\[
  A_2^U\in\Omega^1\left(\left(\overline{\pi}_\zeta\right)^{-1}\left(U\right),\kf\right)
\]
is a $1$-form such that
\[
  \left(p_{K}^{J^1\pi_\zeta}\right)^*A_2^U=\left.A\right|_{\left(\left(\pi_\zeta\right)_1\right)^{-1}\left(U\right)}.
\]
The Lagrangians
\[
  \cL^{\left(1\right)}_{CS}:=h\left(\cL_U\right)
\]
coincide with the local Lagrangians considered in \cite{TejeroPrieto2004}; therefore, we have the following result.
\begin{theorem}\label{thm:Local-to-global}
  Let $M$ be a $\left(2n-1\right)$-dimensional manifold. There exists a (local) one-to one correspondence between the extremals of the variational problem
  \[
    \left(\left[\left(\pi_\zeta\right)_1\right]_1:J^1\left(\pi_\zeta\right)_1\to M,h\left(\lambda_{CS}\right),\mathcal{I}_{\text{con}}\right)
  \]
  restricted to $\left[\left(\pi_\zeta\right)_1\right]_1^{-1}\left(U\right)$ and the variational problem
  \[
    \left(\left(\overline{\pi}_\zeta\right)_1:J^1\overline{\pi}_\zeta\to M,h\left(\cL_U\right),\overline{\mathcal{I}}_{\text{con}}\right).
  \]
\end{theorem}

\begin{remark}
  Another global version for Chern-Simons field theory can be found in \cite{doi:10.1142/S0219887807001990}, where an additional field is introduced, in order to recover the global nature of the transgression form.
\end{remark}
\begin{remark}
  A study of conditions ensuring the existence of global solutions for the local  variational problem for Chern-Simons gauge theory can be found in \cite{Palese2017}. In this regard, it is interesting to note that here we have changed a variational problem determined by local data and whose sections could be globally defined, by a variational problem described by global data, but whose sections are forced to have local nature (part of these sections are sections of a principal bundle).
\end{remark}

\section{Universal variational problem for Chern-Simons gauge theory}
\label{sec:famil-princ-subb}

The present section describes a construction intended to avoid the apparent paradox mentioned in the introduction; as we have said there, the price to pay is to work with a kind of variational problem slightly more general than those usually considered in field theory. Definition \ref{def:griffiths-vars-prob} introduces a handy notation for them.

\par In order to proceed further, let $\pi:P\to M$ be a principal group with structure group $G$. Let $H\subset G$ be a closed Lie subgroup of $G$ and indicate with $R_\zeta\subset P$ a principal subbundle of $P$ associated to the immersion $H\hookrightarrow G$ (the notation will be explained later). Our starting point will be the following diagram
\begin{equation}\label{eq:3dDiagramVariational}
  \begin{tikzcd}[row sep=.8cm,column sep=1cm,ampersand replacement=\&]
    \& J^1\left(\pi_\zeta\right)_1
    \arrow[dl,hook']
    \arrow{rr}{j^1p_H^{J^1\pi_\zeta}}
    \arrow[near start]{dd}{\left(\left(\pi_\zeta\right)_1\right)_{10}}
    \&
    \&
    J^1\overline{\pi}_\zeta
    \arrow[dl,hook']
    \arrow{dd}{\left(\overline{\pi}_\zeta\right)_{10}}
    \\
    J^1\pi_1
    \arrow[crossing over,near start,swap]{rr}{j^1p^P_G}
    \arrow[swap]{dd}{\left(\pi_1\right)_{10}}
    \&
    \&
    J^1\overline{\pi}
    \&
    \\
    \&
    J^1\pi_\zeta
    \arrow[hook']{dl}{}
    \arrow[crossing over,near end]{rr}{p_H^{J^1\pi_\zeta}}
    \arrow[near start]{dd}{\left(\pi_\zeta\right)_{10}}
    \&
    \&
    C\left(R_\zeta\right)
    \arrow[hook']{dl}
    \arrow[bend left=20]{lddd}{\overline{\pi}_\zeta}
    \\
    J^1\pi
    \arrow[crossing over,near start,swap]{rr}{p_G^{J^1\pi}}
    \arrow[swap]{dd}{\pi_{10}}
    \&
    \&
    C\left(P\right)
    \arrow[from=uu,crossing over]
    \arrow{dd}{\overline{\pi}}
    \&
    \\
    \&
    R_\zeta
    \arrow[hook']{dl}{}
    \arrow{dr}{\pi_\zeta}
    \&
    \&
    \\
    P
    \arrow{rr}{\pi}
    \&
    \&
    M
    \&
  \end{tikzcd}
\end{equation}
Recall that $\overline{\pi}_\zeta:C\left(R_\zeta\right)\to M$ is the connection bundle associated to the principal bundle $R_\zeta$.
We will indicate with
\[
  \omega\in\Omega^1\left(J^1\pi,\g\right),\qquad\Omega\in\Omega^2\left(J^1\pi,\g\right)
\]
the canonical connection and its associated curvature form, respectively, on $p_G^{J^1\pi}:J^1\pi\to C\left(P\right)$ and with
\[
  A\in\Omega^1\left(J^1\pi_\zeta,\mathfrak{h}\right),\qquad F\in\Omega^2\left(J^1\pi_\zeta,\mathfrak{h}\right)
\]
the canonical connection and curvature on $p_G^{J^1\pi_\zeta}:J^1\pi_\zeta\to C\left(R_\zeta\right)$. Then, we consider a Lagrangian form $\cL_{CS}\in\Omega^m\left(J^1\pi\right),m=\dim{M},$ fulfilling the following requirements:
\begin{itemize}
\item $\cL_{CS}$ is a polynomial (in the exterior algebra of $J^1\pi$) in $\omega$ and $\Omega$; namely
  \[
    \cL_{CS}=p\left(\omega,\Omega\right).
  \]
  In the particular case of Chern-Simons field theory, we should take $m=3$ and the Lagrangian $3$-form is given by the formula
  \begin{equation}\label{eq:ChernSimonsLagrangian}
    \cL_{CS}:=\left<\omega,\Omega\right>-\frac{1}{6}\left<\omega\stackrel{\wedge}{,}\left[\omega\stackrel{\wedge}{,}\omega\right]\right>.
  \end{equation}
  
\item For $s:U\subset M\to J^1\pi$ a section of $\pi_1:J^1\pi\to M$, $g:U\to G$ a map and
  \[
    \overline{s}:U\to J^1\pi:x\mapsto s\left(x\right)\cdot g\left(x\right),
  \]
  the following transformation property holds
  \[
    \overline{s}^*\cL_{CS}=s^*\cL_{CS}+\mu_s,
  \]
  where $\mu_s\in\Omega^m\left(J^1\pi\right)$ is a closed $m$-form on $J^1\pi$.
\end{itemize}

Accordingly, let us define the Lagrangian $m$-form $\lambda_{CS}$ on $J^1\pi_\zeta$ using pullback
\[
  \lambda_{CS}:=\left(j^1i_\zeta\right)^*\cL_{CS}=p\left(A,F\right).
\]

So far, we have three different variational problems associated to Diagram \eqref{eq:3dDiagramVariational}, which we want to relate, namely
\begin{itemize}
\item The classical variational problem defined by local data on $J^1\overline{\pi}_\zeta$, as described in \cite{TejeroPrieto2004},
\item a global variational problem on $J^1\pi_\zeta$, with action given by the formula
  \[
    s\mapsto\int_Us^*\left(\lambda_{CS}\right),
  \]
  where $s:U\subset M\to J^1\pi_\zeta$ is an arbitrary section of $\left(\pi_\zeta\right)_1:J^1\pi_\zeta\to M$. It is convenient to stress here that sections to be evaluated in the action of this variational problem are arbitrary; namely, we are taking as admissible sections that are not necessarily holonomic. Finally,
\item a Chern-Simons like variational problem on $J^1\pi$: The action is specified by the Lagrangian $m$-form $\cL_{CS}$ and the admissible sections $\sigma:U\subset M\to J^1\pi$ are defined by the requirement
  \[
    \sigma^*\left(\pi_V\circ\omega\right)=0,
  \]
  where $\pi_V:\g\to V$ is the projection associated to a decomposition
  \[
    \g=\mathfrak{h}\oplus V.
  \]
\end{itemize}

The liftings of the variational problems on $J^1\pi$ and $J^1\pi_\zeta$ (as discussed in Section \ref{sec:local-global-chern}) will give us the desired variational problems in the top level of Diagram \eqref{eq:3dDiagramVariational}. In terms of these lifted variational problems we will perform the comparison with the Chern-Simons variational problem on $J^1\overline{\pi}_\zeta$.
So far, we have proved that the extremals of the first and second kind of variational problems can be put in correspondence under mild conditions (Theorem \ref{thm:Local-to-global}); accordingly, the next matter to be addressed is the way in which extremals of the variational problems on $J^1\pi$ and $J^1\pi_\zeta$ are related. To this task is devoted Theorem \ref{thm:univ-chern-simons}: it results that the variational problem on $J^1\pi$ prescribed by the Lagrangian form $\cL_{CS}$ and the set of constraints $\pi_V\circ\omega$ play the role of a kind of \emph{universal variational problem}, whose extremals could be used to represent the extremals of variational problems on any $H$-principal subbundle $R_\zeta\subset P$.

\begin{theorem}\label{thm:univ-chern-simons}
  Let $\cF:=\left\{R_\zeta\right\}$ be a complete family of $H$-subbundles of $P$. Then, for any $H$-principal subbundle $R_\zeta\subset P$, the extremals of $\left(\left(\pi_\zeta\right)_1:J^1\pi_\zeta\to M,\lambda_{CS},0\right)$ are in a local one-to-one correspondence with the extremals of
  \[
    \left(\pi_1:J^1\pi\to M,\cL_{CS},\pi_V\circ\omega\right).
  \]
\end{theorem}
\begin{proof}
  This theorem is consequence of a known fact about connections on subbundles of principal bundles \cite[p. 83]{KN1}: A connection defined on $P$ for a $1$-form $\omega$ is reducible to a connection on $R_\zeta$ if and only if its pullback to $R_\zeta$ is $\mathfrak{h}$-valued.
\par So, let $s:U\to J^1\pi=P\times_M C\left(P\right)$ be an extremal for the variational problem on $J^1\pi$; the fact that
  \[
    s^*\left(\pi_V\circ\omega\right)=0
  \]
  is equivalent to assume that the associated connection $p_G^{J^1\pi}\circ s$ has a connection form $\omega_s$ taking values in $\mathfrak{h}$. Additionally, use the completeness of the family $\cF$ to find a $H$-subbundle $R_\zeta\in\cF$ such that $\left(\pi_{10}\circ s\right)\left(x\right)\in R_\zeta$ for all $x\in U$. Then the fact about connections quoted above will tell us that
  \[
    s\left(x\right)\in R_\zeta\times_M C\left(R_\zeta\right)
  \]
  for all $x\in U$, and also
  \[
    s^*\cL_{CS}=s^*\lambda_{CS}
  \]
  by construction. Now, it could be the case that not any section of $\left(\pi_\zeta\right)_1:J^1\pi_\zeta\to M$ comes through the anterior construction, and so when performing variations on $s$, some of it are lost. In order to prove that this is not so, let $s:U\to J^1\pi_\zeta$ be an extremal of the variational problem $\left(J^1\pi_\zeta,\lambda_{CS},0\right)$. Then
  \[
    \left(\pi_\zeta\right)_{10}\circ s:U\to R_\zeta\subset P
  \]
  is a section of $\pi:P\to M$, and using in this case \cite[Prop. 6.1 in p. 79]{KN1}, we have that the connection
  \[
    p_H^{J^1\pi_\zeta}\circ s:U\to C\left(R_\zeta\right)
  \]
  determines a unique connection $\Gamma_s:U\to C\left(P\right)$ on $P$ with the property that its connection form is $\mathfrak{h}$-valued when pulled back to $R_\zeta$. Therefore, the section
  \[
    \sigma:=\left(\left(\pi_\zeta\right)_{10}\circ s,\Gamma_s\right):U\to J^1\pi
  \]
  verifies that
  \[
    \sigma^*\cL_{CS}=\sigma^*\lambda_{CS}
  \]
  and also $\sigma^*\left(\pi_V\circ\omega\right)=0$, namely, we have lifted a section of $\left(\pi_\zeta\right)_1$ to a section of $\pi_1$. Thus, when performing variations of $s$ as section of $\left(\pi_\zeta\right)_1$, any variation is lifted to a variation of $s$ as section of $\pi_1$; because it is an extremal of the problem
  \[
    \left(\pi_1:J^1\pi\to M,\cL_{CS},\pi_V\circ\omega\right),
  \]
  then $s$ is an extremal of $\left(J^1\pi_\zeta,\lambda_{CS},0\right)$.
  \par Conversely, let us consider an extremal $s_\zeta:U\to J^1\pi_\zeta$ for the variational problem
  \[
    \left(\left(\pi_\zeta\right)_1:J^1\pi_\zeta\to M,\lambda_{CS},0\right);
  \]
  recall from the previous discussion that any such section can be considered as an admissible section of $\pi_1$. Moreover, around every $x\in U$ there exists an open set $U'$, a map
  \[
    g:U'\to G
  \]
  and a section
  \[
    s:U'\to J^1\pi
  \]
  such that
  \[
    s\left(y\right)=s_\zeta\left(y\right)\cdot g\left(y\right)
  \]
  for all $y\in U'$. Thus any variation of $s_\zeta$ as section of $\pi_1$ induces a variation of $s_\zeta$ as section of $\left(\pi_\zeta\right)_1$, for a fixed $\zeta$; by the transformation property of $\cL_{CS}$, as these variations differ in a gauge transformation, $s$ is an extremal for the variational problem on $J^1\pi$ restricted to $U'$.
\end{proof}

\begin{remark}
  Let
  \[
    K:=\left\{A\in GL\left(m,\mR\right):A\eta A^T=\eta\right\}
  \]
  be the Lorentz group associated to the signature matrix $\eta$, as defined in Equation \eqref{eq:LorentzGroup}. When applied to the case
  \[
    P\leftrightsquigarrow AM,\, H\leftrightsquigarrow K\ltimes\mR^3,\, {R_\zeta}\leftrightsquigarrow{O_\zeta}:=\left\{u\in AM:\beta\circ u\text{ is }\zeta-\text{orthogonal}\right\}
  \]
  for any $\zeta:M\to\Sigma:=LM/K$, Theorem \ref{thm:univ-chern-simons} solves the concern (mentioned in the introductory section) underlying the usual descriptions of the correspondence between gravity and Chern-Simons field theory, namely, the fact that it is not clear from the beginning which $\left(K\ltimes\mR^3\right)$-principal bundle should be used, because the metric $\zeta$ is part of the dynamical fields in gravity, and thus it is not fixed in advance. This lack of uniqueness in the choice of the $\left(K\ltimes\mR^3\right)$-bundle disappears when working with the universal variational problem on $J^1\pi$.    
\end{remark}

\begin{remark}
  It could be interested to note that the universality of the variational problem posed by the data
  \begin{equation}\label{eq:VarProbUniversalChernSimons}
    \left(\pi_1:J^1\pi\to M,\cL_{CS},\pi_\pf\circ\omega\right)
  \end{equation}
  is independent of the dimension of the base space $M$. Accordingly, it is suitable for establishing a similar universality result for Chern-Simons Lagrangians intended to describe Lovelock gravity (see for example \cite{Chamseddine1989,Izaurieta:2005vp,Mora2014} and references therein).
\end{remark}

\section{Lift of connections to the affine frame bundle}
\label{sec:lift-conn-affine}

So far we have a universal variational problem \eqref{eq:VarProbUniversalChernSimons} for the family of Chern-Simons gauge theory living in the jet space of a principal bundle $\pi:P\to M$. From now on we will devoted ourselves to particularize this definition to a very specific principal bundle, the so called \emph{affine frame bundle} (see Definition \ref{def:Affine-Frame-Bundle} below), and to relate the variational problem so obtained with Palatini gravity. It results that this relationship is very similar to the very natural relation that there exists between connections on the frame bundle and connections on the affine bundle. To describe this geometrical correspondence is devoted the following section.

We have the splitting short exact sequence
\[
  \begin{tikzcd}[ampersand replacement=\&,row sep=1.5cm,column sep=1.5cm]
    0
    \arrow{r}{}
    \&
    \mathbb{R}^m
    \arrow{r}{\alpha}
    \&
    A\left(m,\mathbb{R}\right)
    \arrow{r}{\beta}
    \&
    GL\left(m,\mathbb{R}\right)
    \arrow[swap]{r}{\beta}
    \arrow[bend left=45,dashed]{l}{\gamma}
    \&
    1
  \end{tikzcd}  
\]
where $A\left(m,\mathbb{R}\right)\subset GL\left(m+1,\mathbb{R}\right)$ is the subgroup of matrices of the form
\[
  B:=
  \begin{bmatrix}
    a&\xi\\
    0&1
  \end{bmatrix}
\]
where $a\in GL\left(m,\mathbb{R}\right)$ and $\xi\in\mR^m$; the maps in the sequence read
\[
  \alpha\left(\xi\right):=\begin{bmatrix}
    1&\xi\\
    0&1
  \end{bmatrix},\qquad
  \beta\left(\begin{bmatrix}
      a&\xi\\
      0&1
    \end{bmatrix}\right):=
  \begin{bmatrix}
    a&0\\
    0&1
  \end{bmatrix},
\]
and
\[
  \gamma\left(a\right):=\begin{bmatrix}
    a&0\\
    0&1
  \end{bmatrix}.
\]
Because this short sequence splits, we can consider
\[
  A\left(m,\mathbb{R}\right)=GL\left(m,\mathbb{R}\right)\oplus\mR^m,
\]
with the isomorphism of groups given by
\[
  GL\left(m,\mathbb{R}\right)\oplus\mR^m\to A\left(m,\mathbb{R}\right):\left(a,\xi\right)\mapsto\gamma\left(a\right)+\alpha\left(\xi\right).
\]
Now, let $A^m$ be the set $\mR^m$ considered as an affine space; we can set an isomorphism between $A\left(m,\mathbb{R}\right)$ and the set of affine maps
\[
  f:A^m\to A^m;
\]
in fact, given $B=\left(a,\xi\right)$, the associated affine map reads
\[
  f_B\left(z\right):=az+\xi
\]
for every $z\in A^m$.

In the same vein, let $A_x\left(M\right)$ be the set $T_xM$ considered as an affine space, for every $x\in M$. As it is well-known \cite{KN1}, the set of affine maps
\[
  u:A^m\to A_x\left(M\right)
\]
for every $x\in M$ has structure of $A\left(m,\mathbb{R}\right)$-principal bundle; the action of an element $B\in A\left(m,\mathbb{R}\right)$ is simply given by 
\[
  u\cdot B:=u\circ B.
\]

\begin{definition}[Bundle of affine frames]\label{def:Affine-Frame-Bundle}
  The \emph{bundle of affine frames on $M$} will be the set
  \[
    AM:=\bigcup_{x\in M}\left\{u:A^m\to A_x\left(M\right)\text{ affine}\right\}.
  \]
\end{definition}

As it follows from the general theory of principal bundles, there exists a pair of principal bundle morphisms associated to the homomorphisms $\beta:A\left(m,\mathbb{R}\right)\to GL\left(m,\mathbb{R}\right)$ and $\gamma:GL\left(m,\mathbb{R}\right)\to A\left(m,\mathbb{R}\right)$
\[
  \begin{tikzcd}[ampersand replacement=\&,row sep=1.5cm,column sep=1.5cm]
    AM
    \arrow[shift right,swap]{rr}{\beta}
    \arrow{dr}{}
    \&
    \&
    LM
    \arrow[shift right,swap]{ll}{\gamma}
    \arrow{dl}{\tau}
    \\
    \&
    M
    \&
  \end{tikzcd}  
\]
For any $G$-principal bundle $\pi:P\to M$, the affine bundle $\left(C\left(P\right),\overline{\pi},M\right)$ defined through the diagram
\[
  \begin{tikzcd}[ampersand replacement=\&,row sep=1.5cm,column sep=1.5cm]
    J^1\pi
    \arrow{r}{\pi_{10}}
    \arrow[swap]{d}{p_G^{J^1\pi}}
    \&
    P
    \arrow{d}{\pi}
    \\
    C\left(P\right):=J^1\pi/G
    \arrow{r}{\overline{\pi}}
    \&
    M
  \end{tikzcd}  
\]
is called the \emph{bundle of connections of the bundle $P$}, and we can establish a canonical one-to-one correspondence between its sections and principal connections on $P$. The correspondence is given as follows: Any element $j_x^1s\in J^1\pi$ is a linear map
\[
  j_x^1s:T_xM\to T_{s\left(x\right)}P
\]
such that
\[
  T_{s\left(x\right)}\pi\circ j_x^1s=\text{id}_{T_xM},
\]
and so a $G$-orbit $\left[j_x^1s\right]_G$ can be interpreted as a linear map
\[
  \left[j_x^1s\right]_G:T_xM\to\left(TP/G\right)_x.
\]
Given $u\in P$, there exists a unique $m$-dimensional subspace $H_u\subset T_uP$ such that
\[
  \left[j_x^1s\right]_G\left(T_xM\right)=p_G^{TP}\left(H_u\right);
\]
the assignment $u\mapsto H_u$ is the connection associated to $\left[j_x^1s\right]_G$.

Therefore we have the diagram
\[
  \begin{tikzcd}[ampersand replacement=\&,row sep=1.5cm,column sep=1.5cm]
    J^1\left(\tau\circ\beta\right)
    \arrow{rr}{p_{A\left(m,\mathbb{R}\right)}^{J^1\left(\tau\circ\beta\right)}}
    \arrow{dr}{j^1\beta}
    \arrow[swap]{dd}{\left(\tau\circ\beta\right)_{10}}
    \&
    \&
    C\left(AM\right)
    \arrow{d}{\left[j^1\beta\right]}
    \\
    \&
    J^1\tau
    \arrow{r}{p_{GL\left(m,\mathbb{R}\right)}^{J^1\tau}}
    \arrow{d}{\tau_{10}}
    \&
    C\left(LM\right)
    \arrow{d}{\overline{\tau}}
    \\
    AM
    \arrow{r}{\beta}
    \&
    LM
    \arrow{r}{\tau}
    \&
    M
  \end{tikzcd}  
\]
It is a theorem that any connection $\Gamma:M\to C\left(LM\right)$ gives rise to a unique connection $\widetilde{\Gamma}:M\to C\left(AM\right)$ such that if
\[
  \omega_0\in\Omega^1\left(LM,\mathfrak{gl}\left(m,\mathbb{R}\right)\right)\qquad\text{ and }\qquad\widetilde{\omega}_0\in\Omega^1\left(AM,\mathfrak{a}\left(m,\mathbb{R}\right)\right)
\]
are the corresponding connection forms, then
\[
\gamma^*\widetilde{\omega}_0=\omega_0+\theta,
\]
where $\theta\in\Omega^1\left(LM,\mathbb{R}^m\right)$ is the canonical solder $1$-form on $LM$.

\begin{definition}[Witten map]
  The \emph{Witten map} $w:\Gamma\left(\overline\tau\right)\to\Gamma\left(\overline{\tau}\circ\left[j^1\beta\right]\right)$ is the map given by
  \[
    w\left(\Gamma\right):=\widetilde{\Gamma}
  \]
  for every connection $\Gamma:M\to C\left(LM\right)$.
\end{definition}

\section{Chern-Simons variational problem with Lie group $A\left(3,\mR\right)$}
\label{sec:chern-simons-vari}

Using Equation \eqref{eq:VarProbUniversalChernSimons} and the geometrical constructions performed in Section \ref{sec:lift-conn-affine}, we will define a Griffiths variational problem on $J^1\left(\tau\circ\beta\right)$ in order to represent Chern-Simons gauge theory. To proceed, let us define the bilinear form $\left<\cdot,\cdot\right>:\mathfrak{gl}\left(m\right)\times\mathfrak{gl}\left(m\right)\to\mR$ given by
\[
  \left<a,b\right>:=\eta^{ij}\eta_{kl}a^k_ib^l_j.
\]
Recall that $K\subset GL\left(m,\mR\right)$ is the Lorentz group defined by the matrix $\eta$ (see Equation \eqref{eq:LorentzGroup}); as always, $\kf$ will indicate its Lie algebra. Then we have the following result.
\begin{lemma}
  The bilinear form $\left<\cdot,\cdot\right>$ is non degenerate and $K$-invariant.
\end{lemma}
Let us now suppose that $m=3$; then we have the isomorphism \cite{Wise_2010}
\[
  \kf\simeq\mR^3
\]
given by
\[
  \xi=\left(\xi^i\right)\mapsto a_i^j:=\eta^{jk}\epsilon_{ikl}\xi^l.
\]
It allows us to use the prescription
\[
  \left<\left(a,\xi\right),\left(b,\zeta\right)\right>:=\left<a,\zeta\right>+\left<b,\xi\right>
\]
for the extension of the bilinear form defined above to $\kf\ltimes\mR^3$; recalling Equation \eqref{eq:Transvections}, we see that it can be extended further to $\mathfrak{a}\left(3\right)$ by declaring that $\pf+0\perp\kf\oplus\mR^3$ and
\[
  \left<\left(a,0\right),\left(b,0\right)\right>=\left<a,b\right>
\]
for any $a,b\in\pf$. Thus we have a quadratic form
\[
  q:\mathfrak{a}\left(3\right)\to\mR:\left(a,\xi\right)\mapsto\left<\left(a,\xi\right),\left(a,\xi\right)\right>.
\]
\begin{definition}[Chern-Simons Lagrangian $3$-form]\label{def:chern-simons-vari}
  The \emph{Lagrangian form for Chern-Simons variational problem} is the $3$-form $\cL_{CS}\in \Omega^3\left(J^1\left(\tau\circ\beta\right)\right)$ defined through
  \[
    \cL_{CS}:=\left<\omega\stackrel{\wedge}{,}\Omega\right>-\frac{1}{6}\left<\omega\stackrel{\wedge}{,}\left[\omega\stackrel{\wedge}{,}\omega\right]\right>,
  \]
  where $\omega\in\Omega^1\left(J^1\left(\tau\circ\beta\right),\mathfrak{a}\left(3\right)\right)$ is the canonical connection on the principal bundle $p_{A\left(3\right)}^{J^1\left(\tau\circ\beta\right)}:J^1\left(\tau\circ\beta\right)\to C\left(AM\right)$.
\end{definition}

Additionally, it is necessary to prescribe the set of constraints that sections $\sigma:U\subset M\to J^1\left(\tau\circ\beta\right)$ should obey in order to be evaluated in the action associated to $\cL_{CS}$. In order to achieve it, let us consider the decomposition
\[
  \mathfrak{a}\left(3\right)=\left(\pf+0\right)\oplus\left(\kf\oplus\mR^3\right);
\]
accordingly, let
\[
  \pi_\pf:\mathfrak{a}\left(3\right)\to\pf+0,\qquad\pi_\kf:\mathfrak{a}\left(3\right)\to\kf\oplus\mR^3
\]
be the corresponding projectors.

\begin{definition}[Constraints for Chern-Simons variational problem on $A\left(3,\mR\right)$]
  We will say that a section $\sigma:U\subset M\to J^1\left(\tau\circ\beta\right)$ is \emph{admissible for the Chern-Simons variational problem} if and only if
  \[
    \sigma^*\left(\pi_\pf\circ\omega\right)=0.
  \]
\end{definition}

With these elements at hand, it is immediate to formulate the variational problem we will use to represent Chern-Simons gauge theory in this context.

\begin{definition}[Chern-Simons variational problem on $A\left(3,\mR\right)$]
  It is the variational problem prescribed by the action
  \[
    \sigma\mapsto\int_U\sigma^*\left(\cL_{CS}\right)
  \]
  for $\sigma:U\subset M\to J^1\left(\tau\circ\beta\right)$ an admissible section.
\end{definition}

In order to fit in the general scheme developed in Section \ref{sec:famil-princ-subb}, it will be necessary to prove that $\cL_{CS}$ established by Definition \ref{def:chern-simons-vari} has the correct transformation properties. To this end is devoted the following lemma, which is a reformulation of a previous result of Freed \cite{Freed:1992vw} to this new setting.

\begin{lemma}
  Let $\left(Q,\pi,N\right)$ be a $H$-principal bundle; indicate with $\omega\in\Omega^1\left(J^1\pi,\hf\right)$ the canonical connection on the bundle
  \[
    p_H^{J^1\pi}:J^1\pi\to C\left(Q\right).
  \]
  Also, let $s:U\subset M\to J^1\pi$ be a local section, and $g:U\to H$ a map. Define the new section
  \[
    \overline{s}:U\to J^1\pi:x\mapsto s\left(x\right)\cdot g\left(x\right).
  \]
  Then
  \[
    \left.\left(\overline{s}^*\omega\right)\right|_x=\text{Ad}_{g\left(x\right)}\circ\left.\left(s^*\omega\right)\right|_x+\left.\left(g^*\lambda\right)\right|_x,
  \]
  where $\lambda\in\Omega^1\left(H,\hf\right)$ is the (left) Maurer-Cartan $1$-form on $H$.
\end{lemma}

\begin{corollary}
  For $\cL_{CS}$ given by Formula \eqref{eq:ChernSimonsLagrangian} and a section $\overline{s}$ constructed as in the previous lemma, the following relation holds
  \[
    \overline{s}^*\cL_{CS}=s^*\cL_{CS}+d\left<\text{Ad}_{g^{-1}}\circ s^*\omega\stackrel{\wedge}{,}g^*\lambda\right>-\frac{1}{6}\left<g^*\lambda\stackrel{\wedge}{,}\left[g^*\lambda\stackrel{\wedge}{,}g^*\lambda\right]\right>.
  \]
\end{corollary}

\begin{remark}
  The $3$-form
  \[
    \left<g^*\lambda\stackrel{\wedge}{,}\left[g^*\lambda\stackrel{\wedge}{,}g^*\lambda\right]\right>
  \]
  is closed; therefore, the last term does not contribute to the action when performing variations, and so
  \[
    \delta\int_M\overline{s}^*\cL_{CS}=\delta\int_Ms^*\cL_{CS}.
  \]
  It means that on $J^1\pi=Q\times_M C\left(Q\right)$, the action associated to $\cL_{CS}$ is independent of the $Q$-factor. Please note also that these considerations can be applied to the variational problem on $J^1\pi_\zeta$ prescribed by the Lagrangian form $\lambda_{CS}$, but cannot to the variational problem on $J^1\pi$, because in the latter case the sections should obey additional constraints, and it reduces the symmetries of the variational problem (namely, not every $g:U\to G$ is allowed, just those taking values in the group of symmetries of the constraints). It tells us that $\cL_{CS}$ should gives rise a variational problem on the reduced bundle
  \[
    J^1\pi/K=Q/K\times_M C\left(Q\right),
  \]
  where $K\subset G$ is the group of symmetries of the constraints.
\end{remark}

Let $\tau_\Sigma:\Sigma\to M$ be the bundle of metrics, obtained from the bundle of affine bundles $AM$ through quotient by the subgroup $K\ltimes\mR^3$, or from the frame bundle $LM$ through quotient by the subgroup $K$,
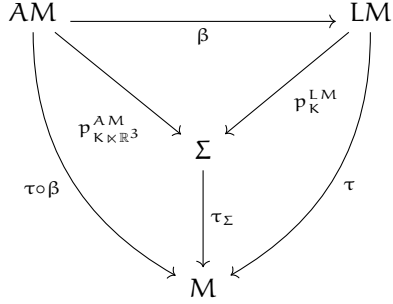
\[
  \begin{tikzcd}[ampersand replacement=\&,row sep=1.3cm,column sep=1.4cm]
    AM
    \arrow[shift right,swap]{rr}{\beta}
    \arrow[swap,near end]{dr}{p_{K\ltimes\mR^3}^{AM}}
    \arrow[bend right=30,swap]{ddr}{\tau\circ\beta}
    \&
    \&
    LM
    \arrow{dl}{p_K^{LM}}
    \arrow[bend left=30]{ddl}{\tau}
    \\
    \&
    \Sigma
    \arrow{d}{\tau_\Sigma}
    \&
    \\
    \&
    M
    \&
  \end{tikzcd}  
\]
The family of $K$-principal subbundles ${O_\zeta}\subset AM$ will be defined as follows. For any section $\zeta:M\to\Sigma$ (that is, a metric on $M$ with $\eta$-signature) we define
\[
  {O_\zeta}:=\left\{u\in AM:\beta\circ u\text{ is }\zeta-\text{orthogonal}\right\}.
\]
Because any section $s=\left(v;X_1,\cdots,X_m\right):U\to AM$ gives rise to a metric according to the formula
\[
  \zeta_s:=\eta^{ij}X_i\otimes X_j,
\]
it follows that
\[
s\left(x\right)\in{O_{\zeta_s}}\qquad\text{for all }x\in U,
\]
and this family of subbundles will be complete (in the sense of Definition \ref{def:complete-family}). Thus, we are in the scope of Theorem \ref{thm:univ-chern-simons}; accordingly, the Chern-Simons variational problem and the variational problem defined on $J^1\pi_\zeta$ have a one-to-one correspondence between its extremals, for any metric $\zeta$.

\section{Chern-Simons gauge theory for $A\left(3,\mR\right)$ and gravity}
\label{sec:gauge-theory-gravity}

In the present section we will give an interpretation to the universal variational problem for Chern-Simons theory: It will become the variational problem for Palatini gravity, but viewed from the viewpoint of affine geometry. Namely, using Diagram \eqref{eq:3dDiagramVariational} adapted to this particular case, we have the following chain of maps
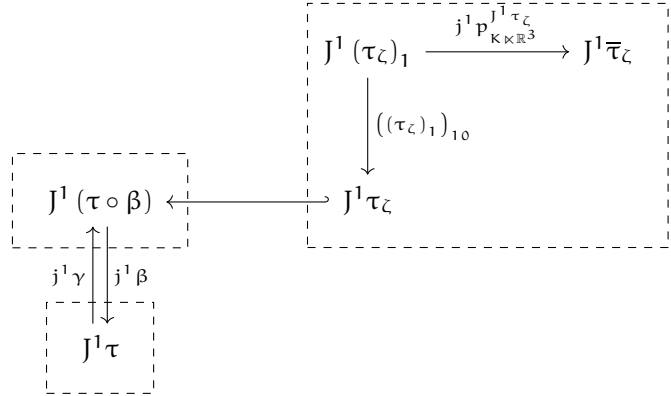
\[
  \begin{tikzcd}[remember picture,ampersand replacement=\&,row sep=1.3cm,column sep=1.9cm,
     execute at end picture={
       \node[dashellipse=(\tikzcdmatrixname-3-1)(\tikzcdmatrixname-3-1)]{};
       \node[dashellipse=(\tikzcdmatrixname-2-2)(\tikzcdmatrixname-1-3)]{};
       \node[dashellipse=(\tikzcdmatrixname-2-1)(\tikzcdmatrixname-2-1)]{};
     }]
    \&
    J^1\left(\tau_\zeta\right)_1
    \arrow{r}{j^1p_{K\ltimes\mR^3}^{J^1\tau_\zeta}}
    \arrow{d}{\left(\left(\tau_\zeta\right)_1\right)_{10}}
    \&
    J^1\overline{\tau}_\zeta
    \\
    J^1\left(\tau\circ\beta\right)
    \arrow[shift left]{d}{j^1\beta}
    \&
    J^1\tau_\zeta
    \arrow[hook']{l}{}
    \&
    \\
    J^1\tau
    \arrow[shift left]{u}{j^1\gamma}
    \&
    \&
  \end{tikzcd}
\]
for $\zeta:M\to\Sigma$ a metric with $\eta$-signature. The upper right box encircles bundles whose structure group is $K\ltimes\mR^3$; the upper left box contains the bundle with structure group $GL\left(3,\mR\right)\ltimes\mR^3$, and finally, the lower dashed rectangle indicates the bundle whose structure group is $GL\left(3,\mR\right)$. So far we have proved that local variational problems on $J^1\overline{\tau}_\zeta$ can be globalized to a family of variational problems on $J^1\tau_\zeta$; moreover, we have shown how to represent this family of variational problems with a unique variational problem of Griffiths type on $J^1\left(\tau\circ\beta\right)$. What we will show in the present section is how the canonical maps between the frame bundle and the affine frame bundle allow us to interpret this universal variational problem as the affine version of the Palatini variational problem.

\par Using the canonical basis $\left\{e_i\right\}$ for $\mR^3$ and $\left\{E_i^j\right\}$ for $\mathfrak{gl}\left(3,\mR\right)$, we have
\[
  \overline{\theta}=\overline{\theta}^ie_i,\qquad\overline{\omega}=\overline{\omega}^i_jE^j_i,\qquad\overline{\Omega}=\overline{\Omega}^i_jE^j_i
\]
for the canonical forms on $J^1\tau$.

Also, let us recall \cite{doi:10.1142/S0219887818500445,Capriotti:2019bqh} that the following $3$-form on $J^1\tau$,
\[
  \lambda_{PG}:=\eta^{kl}\epsilon_{lij}\overline{\theta}^i\wedge\overline{\Omega}^j_k,
\]
is a Lagrangian form for Palatini gravity; in particular, it can be written
\[
  \lambda_{PG}=\left<\overline{\theta}\stackrel{\wedge}{,}\overline{\Omega}\right>=\left<\overline{\theta}\stackrel{\wedge}{,}d\overline{\omega}+\frac{1}{2}\left[\overline{\omega}\stackrel{\wedge}{,}\overline{\omega}\right]\right>.
\]
In any case, it gives rise to a variational problem
\[
  \left(\tau_1:J^1\tau\to M,\lambda_{PG},\pi_\pf\circ\overline{\omega}\right)
\]
whose extremals are solution of the (vacuum) $3$-dimensional Einstein equations; in this context, the constraints $\pi_\pf\circ\omega$ are the so called \emph{metricity conditions}, ensuring that if a section
\[
  s:U\to J^1\pi
\]
is admissible, the parallel transport defined by the connection
\[
  p_K^{J^1\tau}\circ s:U\to C\left(LM\right)
\]
is an isometry for the metric
\[
  g=\eta^{ij}X_i\otimes X_j,
\]
where
\[
  \tau_{10}\left(s\left(x\right)\right)=\left(X_1\left(x\right),\cdots,X_m\left(x\right)\right)\in L_xM
\]
for all $x\in U$.

\par On the other side, we can write down
\begin{align*}
  \cL_{CS}&=\left<\omega\stackrel{\wedge}{,}\Omega\right>-\frac{1}{6}\left<\omega\stackrel{\wedge}{,}\left[\omega\stackrel{\wedge}{,}\omega\right]\right>\\
  &=\left<\omega\stackrel{\wedge}{,}\left(d\omega+\frac{1}{3}\left[\omega\stackrel{\wedge}{,}\omega\right]\right)\right>.
\end{align*}
Let $\sigma:U\to J^1\tau$ be a local section of the jet space of the frame bundle, and define
\[
  \widetilde{s}:=\gamma\circ\pi_{10}\circ\sigma:U\to AM;
\]
also, define $\widetilde{\Gamma}:U\to C\left(AM\right)$ using the Witten map on the connection $\Gamma:=p_{GL\left(m\right)}^{J^1\tau}\circ\sigma$; then we have a section
\[
  \widetilde{\sigma}:U\to J^1\left(\tau\circ\beta\right):x\mapsto\left(\widetilde{s}\left(x\right),\widetilde{\Gamma}\left(x\right)\right).
\]
In this setting the following result holds; it is the classical result \cite{MR974271,Wise_2010} written in terms of the constructions made in this context.
\begin{theorem}
  In the setting described above, the following identity
  \[
    \sigma^*\left(\lambda_{PG}\right)=\widetilde{\sigma}^*\left(\cL_{CS}\right)
  \]
  holds.
\end{theorem}

\begin{corollary}\label{thm:chern-simons-palatini}
  The extremals of the variational problems
  \[
    \left(\tau_1:J^1\tau\to M,\lambda_{PG},\pi_\pf\circ\overline{\omega}\right)
  \]
  and
  \[
    \left(\left(\tau\circ\beta\right)_1:J^1\left(\tau\circ\beta\right)\to M,\cL_{CS},\pi_\pf\circ{\omega}\right)
  \]
  are in one-to-one correspondence through the maps $\beta$ and $\gamma$.
\end{corollary}

\section{Conclusions and outlook}
\label{sec:conclusions}

In the present article an interpretation of the correspondence between gravity and Chern-Simons field theory on three dimensional spacetime is given in terms of the geometrical correspondence between the frame bundle and its affine counterpart. Although in some sense this equivalence is reached by the theory developed by Wise \cite{wise2009symmetric}, our work departed from it in two ways:
\begin{itemize}
\item First, the variational problems discussed here live on the jet space of a principal bundle, and
\item we have given an interpretation of the affine counterpart of the Palatini gravity Lagrangian, as the Lagrangian for an universal variational problem describing the full family of geometrical Chern-Simons variational problem.
\end{itemize}
An extension of this scheme to the constructions in terms of Cartan geometry considered by Wise in the previously cited article will be carried out elsewhere.

\printbibliography

\end{document}